\documentclass[twocolumn, doublespace, 12pt fonts]{aastex6}
\usepackage{amsmath}
 \usepackage{lineno}

\shorttitle{A simple model for radio pulsar nulling}
\shortauthors{DS}

\date{\today}
\begin{document}

\title{On the transition stage of pulsar pulsed radio emission and its potential association with radio pulsar nulling}

\author{Shuang Du$^{1}$}
\affil{
$^{1}${School of Mathematics and Computer, Tongling University, Tongling, Anhui, 244061, China}}

\email{dushuang@pku.edu.cn}

\begin{abstract}
While the precise mechanism of generating pulsed coherent radio emission from pulsars remains elusive,
certain gap-invoking models (especially, the inner gap model) offer a comprehensive and plausible explanation for the genesis and termination of such emissions.
However, the transition stage between the period of persistent radio emission and the period of radio quiet remains poorly understood,
despite observations indicating that a radio pulsar in the pulse nulling state is undergoing the transition stage.
In this study, we present a qualitative explanation for the elusive transition stage by modeling pulsar magnetospheres analytically as equivalent $RC$ circuits based on the inner gap model.
Our result indicates that, due to lengthy spin-down, older radio pulsars will gradually shift from the state of persistent radio emission to a certain type of pulse nulling state by delayed sparks within their inner gaps. 
\end{abstract}


\section{Introduction}\label{sec1}
The pulsar death line serves as the dividing line for the evolutionary process of radio pulsars,
signifying that pulsed radio emission will cease when specific pulsar parameters evolve to certain values \citep{1975ApJ...196...51R,1993ApJ...402..264C,2000ApJ...531L.135Z}.
The states of ongoing radio emission and radio quiet can be effectively distinguished through observations.
However, the transition stage between these two states remains unclear and necessitates further exploration
though observations imply that pulsars in the pulse nulling state may be pulsars in this transition stage.

Pulsar nulling is the sudden disappearance of pulsed radio emission of pulsars \citep{1970Natur.228...42B}.
Statistical analyses \citep{1976MNRAS.176..249R,2007MNRAS.377.1383W} have revealed a correlation between the null fraction (the fraction of time a pulsar spends in a null state) and pulsar age (or period; see \citealt{1992ApJ...394..574B}), with older pulsars more likely to null.
This statistical result suggests that nulling phenomena may be related to certain gap-invoking models of pulsar pulsed radio emission \citep{1975ApJ...196...51R,1979ApJ...231..854A},
as these models predict that the acceleration of charged particles responsible for radio radiation weakens as the increases of pulsar ages and periods.
Hence, nulling phenomena can be also related to the transition stage since both of the two are the results of the decrement of discharge capability of pulsars under the gap-invoking models.
However, the recent statistical analysis \citep{2021MNRAS.502.4669S} shows a weak correlation between nulling fraction and spin period which could be attributed to selection
effects.
On the other hand, these is another possibility that pulsar nulling may be a kind of mode changing of the pulsed radio emission \citep{2005MNRAS.356...59E}.
Some authors have proposed that geometric effects of pulsar radio beams could induce nulling phenomena.
\cite{2010MNRAS.408L..41T} assume that a pulsar magnetosphere has different states with different cross sections of the corotating zone.
The switches between these states result in the radio beam alternately passing the line of sight. When the radio beam is out of the line of sight, the pulse is null.
\cite{2021A&A...653L...3D} proposed a more physical model in which the invisible radio beam are caused by the wobble of the radio beam induced by the $\mathbf{E}\times \mathbf{B} $ drift \citep{1975ApJ...196...51R}.

So even if the implication of the above statistical results \citep{1976MNRAS.176..249R,1992ApJ...394..574B,2007MNRAS.377.1383W} is correct,
a physical and analytical explanation is still needed just like that of the geometric models \citep{2010MNRAS.408L..41T, 2021A&A...653L...3D}.
However, despite their capability to explain critical issues like the acceleration of charge particles necessary for generating radio emissions and the pulsar death line, these gap-invoking models still lack clarity on the matter of whether pulsed radio emissions fade gradually as the associated pulsar parameters evolve, or abruptly end upon reaching certain values of these parameters.
Given the superiorities of the inner gap model \citep{1975ApJ...196...51R}, for example it offers a satisfactory explanation for drifting pulses,
if this model accurately describes the physics of pulsar radio emissions,
it should also be capable of predicting the existence of the transition stage as well as some corresponding phenomena (e.g., pulse nulling).

In Section \ref{sec2}, we model pulsar magnetospheres equivalently as a resistive-capacitive ($RC$) circuit based on the inner gap model.
In Section \ref{sec3}, through this equivalent model, we present a qualitative discussion on the missing component in the evolution of pulsed radio emission.
Lastly, we would like to clarify that pulsars in the transition stage are characterized by delayed sparks, and they can naturally manifest as a kind of nulling pulsars.

\section{equivalent $RC$ circuit model}\label{sec2}

Let us first briefly review the inner gap model.
For a pulsar with net positive charges being in its open field line region,
when these positive charges move outwards \citep{1971ApJ...164..529S,1973NPhS..246....6H},
the lost charges can not be replenished by the stellar surface due to large binding energy of positive charges.
Whereafter, a gap grows on the polar cap.
The potential across the gap, $U$, increases with the gap height, $h$, initially.
When the voltage increases to a critical value (inversely proportional to the curvature radius of magnetic field lines),
positrons (e.g., originally produced by the thermal $\gamma$-ray photons from the pulsar surface through $\gamma \mathbf{B}$ processes
 that $\gamma+\mathbf{B}\rightarrow e^{-}+e^{+}+\mathbf{B}$)
in the gap can be accelerated to a high energy that the photons emitted by these positrons via curvature radiation and even inverse Compton scattering \citep{1998A&A...333..172Q}
will again be converted into electron-positron pairs via $\gamma \mathbf{B}$ processes.
Avalanches of discharges lead to the eventual reduction of the potential $U$, and after which the spark ceases.
The growth and reduction of the gap happens back and forth, generating a large number of secondary charges to sustain the radio emission.
However, the maximum value of the potential across the gap does not grow endlessly with the gap height,
but is also enslaved by the pulsar period (e.g., the potential is inversely proportional to the pulsar period).
As the pulsar spins down, there comes a point where the most powerful potential within the inner gap can no longer sustain a spark,
leading to the ultimate extinction of pulsed radio emission.

We note that, during the charge process (the growth of the gap height),
the electric field across the gap behaves like that of a parallel-plate capacitor since the gap height is much smaller than the radius of the polar cap, $r_{\rm p}$.
The electric field of a parallel-plate capacitor is given by
\begin{eqnarray}
{E}'=\frac{{U}'}{{d'}}=\frac{4\pi k Q'}{S'}
\end{eqnarray}
where $U'$ and $d'$ are the voltage and interval between the two plates, $Q'$ and $S'$ are the charge and area of each plate, and $k$ is the electrostatic constant.
As a contrast, the electric field across the gap is \citep{1975ApJ...196...51R}
\begin{eqnarray}\label{e2}
E\approx \frac{2U}{h} =\frac{2\Omega B_{\rm s}}{c}h,
\end{eqnarray}
where $\Omega$ is the spin velocity, and $B_{\rm s}$ is the magnetic field strength on the polar cap.
Therefore, we equivalently treat the inner gap as a parallel-plate capacitor with the voltage being $U'=U$ and the area of the plate, $S'$, being the area of the polar cap, $S$.
Then, the equivalent interval is $h/2$,
the equivalent charge is
\begin{eqnarray}
Q=\frac{\Omega B_{s}hS}{2\pi ck},
\end{eqnarray}
and the equivalent capacitance is
\begin{eqnarray}\label{a0}
C=\frac{Q}{U}=\frac{S\varepsilon}{2\pi k h},
\end{eqnarray}
where $\varepsilon$ is the permittivity.
Correspondingly, the increase rate of the gap height is exactly the bulk velocity of the charge particles which are flowing outwards.

Now, the charge and discharge in the inner gap can be equivalent to the charge and break down of the capacitor.
Such an equality is nonlinear in the sense that both of the capacitance and charge increase with the gap height until the capacitor is broken down.
The governing equation of the charge of the pulsar-magnetosphere system is
\begin{eqnarray}\label{a1}
R\frac{ d(C U)}{dt}+U=U_{\rm max},
\end{eqnarray}
where $R$ is the equivalent resistance of the circuit (which should be  approximately a constant since the whole magnetosphere is in a quasi-static state during a short duration, as well as the spin velocity $\Omega$),
$U=U(t)$ is the potential across the gap at time $t$ which should be given by  (\citealt{1975ApJ...196...51R};
this means that the charge in the inner gap is a transient process as with the charge of a realistic $RC$ circuit)
\begin{eqnarray}\label{a3}
U\approx\frac{\Omega B_{\rm s}}{c}h(t)^{2},
\end{eqnarray}
and
\begin{eqnarray}\label{a4}
U_{\rm max}\approx\frac{\Omega B_{\rm s}}{c}h_{\rm max}^{2}
\end{eqnarray}
is the supply voltage (e.g., due to the unipolar induction) with $h_{\rm max}$ being the maximum thickness of the inner gap.
It is worth reminding that gap voltage, $U$, may not be able to reach the supply voltage, $U_{\rm max}$, since before that time the capacitor may have been broken down.
As illustrated in the inner gap model, the maximum possible potential drop along
any magnetic field line within the polar cap is
\begin{eqnarray}
\Delta V_{\rm max}\approx \frac{\Omega B_{\rm s}}{c} \frac{r_{\rm p}^{2}}{2},
\end{eqnarray}
that is $h_{\rm max}=0.7 r_{\rm p}$.
However, this potential drop is usually larger than the voltage, $U_{\rm sp}$, required for a spark (see equation (23) in {\citealt{1975ApJ...196...51R}}).
Therefore, when the value of $U$ reaches $U_{\rm sp}$, the growth of the gap height will stop.

Note that $S\approx 4\pi r_{\rm p}^{2}$ and
\begin{eqnarray}
r_{\rm p}=r_{\ast}\left ( \frac{\Omega r_{\ast}}{c} \right )^{1/2},
\end{eqnarray}
where $r_{\ast}$ is the radius of the pulsar, equations (\ref{a0})-(\ref{a4}) can be solved as (Riccati equation of $h$)
\begin{eqnarray}\label{a7}
h(t)=\frac{1}{\alpha}\frac{c_{1}\sqrt{\alpha\beta}e^{\sqrt{\alpha\beta}t}-c_{2}\sqrt{\alpha\beta}e^{-\sqrt{\alpha\beta}t}}{c_{1}e^{\sqrt{\alpha\beta}t}+c_{2}e^{-\sqrt{\alpha\beta}t}},
\end{eqnarray}
where
\begin{eqnarray}\label{e11}
\alpha=\frac{kc}{2\Omega r_{\ast}^{3}R\varepsilon}, \;\;\;\; \beta=\frac{kc}{2\Omega r_{\ast}^{3}R\varepsilon} h_{\rm max}^{2},
\end{eqnarray}
and $c_{1}$, $c_{2}$ are integration constants.
We set the initial condition as $h(t=0)=0$, and have $c_{1}=c_{2}$ through equation (\ref{a7}). Therefore, the gap hight is given by
\begin{eqnarray}\label{e13}
h(t)=h_{\rm max}\tanh (\sqrt{\alpha\beta}t).
\end{eqnarray}
Through equation (\ref{e13}), the bulk velocity of charge outflow is
\begin{eqnarray}\label{e14}
v_{\rm of}=\frac{dh(t)}{dt}=h_{\rm max}\sqrt{\alpha\beta}[1-\tanh^{2} (\sqrt{\alpha\beta}t)].
\end{eqnarray}
According to equations (\ref{a3}) and (\ref{e13}), we get
\begin{eqnarray}
U=\frac{\Omega B_{\rm s}}{c}[h_{\rm max}\tanh (\sqrt{\alpha\beta}t)]^{2},
\end{eqnarray}
and
\begin{eqnarray}\label{e15}
U_{\rm sp}=\frac{\Omega B_{\rm s}}{c}[h_{\rm max}\tanh (\sqrt{\alpha\beta}\tau)]^{2},
\end{eqnarray}
where $\tau$ is the time that the potential across the gap increases to trigger a spark (then the spark frequency is $\sim 1/\tau$; but this is an ideal case).

The voltage required for a spark in the inner gap, $U_{\rm sp}$, is usually a constant in a long term as long as the magnetic field is stable (but when $U_{\rm max}$ gets close to $U_{\rm sp}$, the disturbance of $U_{\rm sp}$ may have evident effects due to the asymptotic property of hyperbolic tangent functions; see Section \ref{sec3}).
Then, equation (\ref{e15}) shows that the time it takes for a spark, $\tau$, will increase with the spindown of the pulsar.
This is not the only effect of pulsar spin-down.
Due to the asymptotic property of hyperbolic tangent functions, this dependence between $U_{\rm sp}$ and $\tau$ is coincidently corresponding to a certain type of pulsar nulling phenomena.

\section{Explanation of pulse nulling}\label{sec3}
\begin{figure}[htbp]
\centering
\includegraphics[scale=.35]{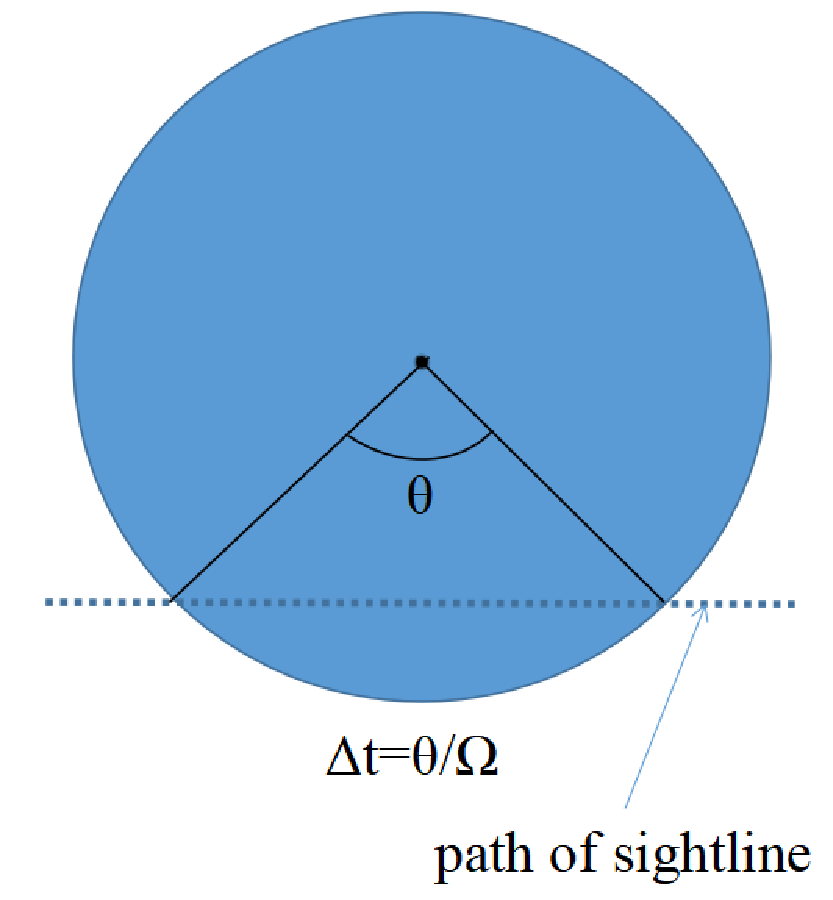}
\caption{The geometry between the radio beam and the line of sight. The blue circle is the cross section swept by the radio beam.
}
\label{f1}
\end{figure}

\begin{figure}[htbp]
\centering
\includegraphics[scale=.34]{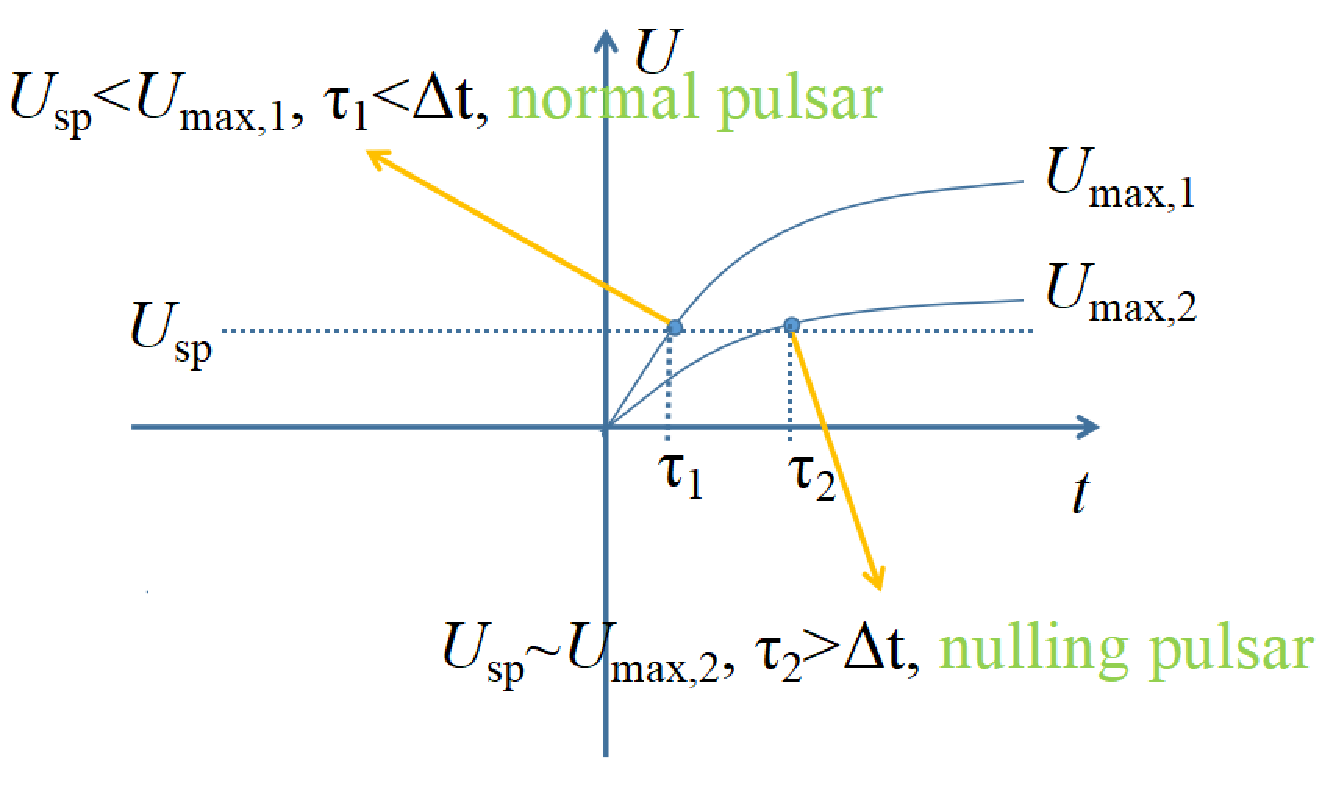}
\caption{An explanation of pulse nulling.
The two thin solid curves are the sketch of the evolution of the potential across the gap versus the time.
}
\label{f3}
\end{figure}

In addition to the geometrical explanations introduced above \citep{2010MNRAS.408L..41T,2021A&A...653L...3D},
according to the geometry shown in Figure \ref{f1}, when the radio beam crosses the line of sight,
if no spark is generated in the inner gap during this interval $\Delta t$, the pulsar will experience a null state during this period.
Equation (\ref{e15}) simply supplies the necessary condition for a delayed spark.

As shown in Figure \ref{f3}, the two thin solid curves are the sketch of the evolution of the potential across the gap versus the time.
The difference between the two curves is that they have different supply voltages that $U_{\rm max,1}> U_{\rm max,2}$.
When the pulsar is young, the spin-down of the pulsar is inadequate and the value of the supply voltage, $U_{\rm max,1}$, is large enough that $U_{\rm sp}<U_{\rm max,1}$.
Under this situation, the time it takes for a spark, $\tau_{1}$, can be shorter than the duration $\Delta t$,
so the pulsar can produce a radio pulse during the the sightline passage time for each period and acts as a normal radio pulsar.
As the spin-down of the pulsar, the supply voltage will decay to a certain value, $U_{\rm max,2}$ (e.g., $U_{\rm max,2} \sim U_{\rm sp}$).
With the decrease of the supply voltage that from $U_{\rm max,1}$ to $U_{\rm max,2}$, the charging time for a spark becomes significantly longer, that is from $\tau_{1}$ to $\tau_{2}$.
Due to the asymptotic property of hyperbolic tangent functions, as long as the pulsar is old enough (i.e., the value of $P$ is large enough),
there will be $\tau_{2}>\Delta t$, finally. Under this critical condition, since the potential across the gap do not have sufficient time to increase to the value required for a spark in the duration $\Delta t$,
the pulsar begins to null. Another characteristic situation is $\tau_{2}>P$ under which the pulsar must null after a successful pulse.

It is worth noting that, although a significant positive correlation exists between nulling fraction and pulsar period under this equivalent model (nulling is inevitable as the spin-down of the pulsar),
this correlation could be influenced by other parameters such as $\theta$ and $\Omega$.
For instance, pulse nulling could be observed in young pulsars if the value of $\theta$ is sufficiently small and the value of $\Omega$ is large enough.
On the other hand, after each spark, the voltage of the gap does not directly drop to zero, which shortens the charging time, $\tau$, required for the next spark.
The true charging time of a spark is the time that the gap voltage increases from a certain value (the gap voltage after the last discharge) to the voltage need by this spark.
Besides, the value of the voltage required for each spark, $U_{\rm sp}$, may not be a constant due to the disturbance from the previous spark.
These random modulations will significantly impact the charging time of the subsequent spark and the off-state length.
These supplementary effects could disrupt the positive correlation between nulling fraction and pulsar age (period) and
lead to the weak correlation between nulling fraction and spin period shown in \cite{2021MNRAS.502.4669S}.

At all events, under this equivalent model, between the epoch of persistent radio emission ($U_{\rm sp}<U_{\rm max}$, $\tau<\Delta t$) and the epoch of radio quiet ($U_{\rm max}<U_{\rm sp}$),
there should be indeed a transitional pulse nulling stage ($U_{\rm sp}\sim U_{\rm max}$, $\tau>\Delta t$).

\section{Summary and discussion}
In this paper, we present an equivalent $RC$ circuit to model the charge of pulsar inner gaps according to the inner gap model.
We show that the charging time of sparks in pulsar inner gaps is influenced by the spin-down of pulsars.
As pulsars spin down, the charging time of sparks increases, ultimately leading to a kind of nulling phenomena due to these delayed sparks.
This phenomenon is inevitable and can serve as an indicator that the pulsar is in the transition stage of its pulsed radio emission.

Although we discuss in Section \ref{sec3} that the positive correlation between nulling fraction and pulsar age (period) could be disrupted by other effects,
it is important to note that the result shown in \cite{2021MNRAS.502.4669S} could also be attributed to multiple origins of nulling phenomena.
For instance, the nulling phenomenon could be categorized into two types: nulling with mode changing (e.g., \citealt{2005MNRAS.356...59E}) and nulling without mode changing. It is challenging to comprehend the nulling pulsar with mode changing under our current model, as under this equivalent model the radio emission should completely disappear during the nulling state, rather than remaining at a very low intensity.
To address this issue under our model, we have to assume that, in addition to inner gaps, there exist additional mechanisms that accelerate charged particles,
such as ion outflows that result in deviations from the Goldreich-Julian density \citep{1979ApJ...231..854A}\footnote{see \cite{HCK} and \cite{KLO} for other discussions}.
Moreover, the longer the charging time, the weaker the ability of continuous sparks.
Consequently, it is difficult to comprehend the pulsar with a significant nulling fraction and long-lasting continuous pulsed radio emission under our equivalent model.
The extreme case of a pulsar with a large nulling fraction and alternating between a single pulse and a long off-state state aligns well with our model.

If the inner gap model can be further verified by future observations,
the equivalent model shown in this paper has the potential to contribute to a better understanding of the mechanism behind the generation of pulsar coherent radio emissions.
Specifically, the result that there could be $\tau>\Delta t$ predicted by this equivalent model supports the scenario in which coherent pulsed radio emissions of pulsars are generated by charged bunches (e.g., \citealt{1975ApJ...196...51R,1987ApJ...320..333U}).
Meanwhile, $\tau>\Delta t$ does not support the scenario in which the coherent pulsed radio emissions are generated by collective oscillations of charged particles \citep{2008ApJ...683L..41B,2020PhRvL.124x5101P}.

\section{Acknowledgement}
We would like to thank the anonymous reviewer for the useful comments.
This work is supported by a research start-up grant from Tongling University (2023tlxyrc14).


\end{document}